\documentclass[preprint,prb,showpacs,showkeys,superscriptaddress,titlepage]{revtex4}
\usepackage{amsfonts}
\usepackage{amsmath}
\usepackage{amssymb}
\usepackage{graphicx}

\begin{document}

\title{Hybrid expansions for local structural relaxations}
\author{H. Y. Geng}
\affiliation{Laboratory for Shock Wave and Detonation Physics
Research, Southwest Institute of Fluid Physics, P. O. Box 919-102,
Mianyang Sichuan 621900, China}
\affiliation{Department of Physics, Tsinghua University, Beijing 100084, China}
\author{M. H. F. Sluiter}
\altaffiliation{Institute for Materials Research, Tohoku University, Sendai, 980-8577 Japan}
\affiliation{Laboratory of Materials Science, Delft University of Technology, 2628AL Delft, the Netherlands}
\author{N. X. Chen}
\altaffiliation{Institute for Applied Physics,
University of Science and Technology, Beijing 100083, China}
\affiliation{Department of Physics, Tsinghua University, Beijing
100084, China}

\begin{abstract}
A model is constructed in which pair potentials are combined with the cluster expansion method in order to better describe the energetics of structurally relaxed substitutional alloys.
The effect of structural relaxations away from the ideal crystal positions, and the effect of ordering is described by interatomic-distance dependent pair potentials, while more subtle configurational aspects associated with correlations of three- and more sites are described purely within the cluster expansion formalism.
Implementation of such a hybrid expansion in the context of the cluster variation method or Monte Carlo method gives improved ability to model phase stability in alloys from first-principles.
\end{abstract}

\volumeyear{year}
\volumenumber{number}
\issuenumber{number}
\eid{identifier}
\maketitle


The lattice gas model has been very effective for modeling substitutional and interstitial alloys
and compounds.\cite{ducastelle91}
Although it is a generalization of the classical Ising model, when combined with effective interactions extracted from  \emph{ab initio} total energies through the cluster expansion method (CEM),\cite{connolly83} it provides the basic framework for the modern theory of alloys.\cite{ducastelle91,defontaine94,vdwalle02}

Experience has learned that on perfect lattices the CEM converges rapidly, requiring only clusters with a few sites for the thermodynamic modeling of alloy phase stability, see e.g. Ref.~[\onlinecite{sluiter96}].
However, when structural relaxations play an important role, as in alloys involving constituents with large size differences, convergence of the CEM becomes poor and typically long-ranged effective pair and many-body interactions are necessary.\cite{luzw91}
In some alloy systems relaxations effects are dominant.\cite{colin00}
In previous calculations this problem was treated by fully relaxing structures when performing first-principles calculations and performing the cluster expansion not over the internal energy but over other expedient thermodynamic potentials such as the enthalpy.\cite{sluiter01}
However, this method fails in a number of instances:
(\romannumeral 1) When the terminal phases have different crystal structures, it frequently happens that relaxation of the unstable structures leads to the stable structure.
For example in the case of bcc and fcc structures, while intermediate relaxed structures may exist in clearly fcc or bcc derived form, for the pure endpoints it has been found that when either fcc or bcc is stable, unrestricted relaxation of the unstable structure is not possible without arriving at the stable structure.
(\romannumeral 2) Another difficulty is that as a function of temperature, quite apart from configurational changes, there are also changes in the relaxed structure as a result of the lattice vibrations.~\cite{asta96}
This aspect of the temperature dependence of the relaxation energy is not presented in the current implementation of the CEM.
(\romannumeral 3) When structures relax it can become impossible to uniquely associate a relaxed cluster with a cluster in the unrelaxed structure.
For example, an fcc-based ordered structure such as L1$_0$ that relaxes to the bcc-based B2 structure presents topological difficulties in that it becomes very difficult to properly categorize and count the nearest- and second- nearest neighbor pairs.
Although for this specific problem an intermediate body centered tetragonal structure can be devised that describes both fcc and bcc crystal structures,~\cite{turchi} generally, when several modes of distortion exists the definition of intermediate or generalized underlying crystal structures becomes impractical.
These limitations of the CEM come about because it was originally conceived for fixed perfect lattices and not for relaxed structures.\cite{connolly83}
Current implementations of the CEM give errors that typically are of the order of 300 K for order-disorder transition temperatures $T_{c}$.
Kikuchi claimed that lattice distortion is the main reason for the discrepancy between \emph{ab initio}-based and  experimental $T_{c}$.\cite{kiku99-02}
As an illustration one can mention the $T_{c}$ of $L1_{2}$ Ni$_{3}$Al: when unrelaxed structures are used in the CEM, a fortuitous agreement with experiment is found.\cite{pasturel92,geng04}
When relaxed structures are used in CEM the agreement with experiment worsens\cite{geng04} and only a better description of structural relaxations can improve agreement, see e.g. Refs.~[\onlinecite{asta96}] and [\onlinecite{sahara04}].
For simulations of kinetics also, relaxation from the ideal lattice sites may greatly influence the energy barriers for atom-exchanges.
Therefore, current kinetic Monte Carlo diffusion modeling with local-environment independent parameters might be made more realistic by explicitly treating relaxations.
A proper accounting for local distortions might much improve the change of the melting temperature under high pressure also.\cite{geng05}

Here, it is our aim to treat the effects of relaxations by proposing an efficient energy functional.
Firstly, an energy function that depends on atomic position must be constructed, which is completely different from the usual CEM scheme.
Obviously, a pair potential model provides the simplest and most widely used function with this characteristic.
Although pair contributions are the largest energy contribution in alloys and compounds, effective multi-body interactions are necessary for accurately describing cluster occupation competitions.\cite{ducastelle91,hoshino04}
However, atomic position dependent multi-body potentials are difficult to obtain and implement, so that we opt to include these effects through the effective cluster interactions (ECI) as efficiently generated by the CEM.
Clearly, it is attractive to combine the CEM and the pair potential approach to model the lattice distortion energy.

\begin{figure}[!htbp]
\centering
\includegraphics[width=5cm]{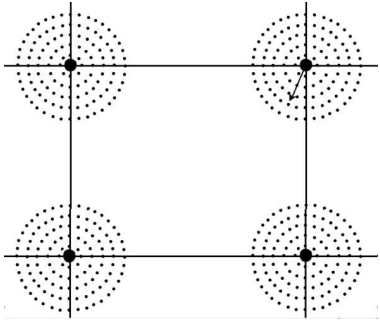}
\caption{Distortion model for the 2-dimensional square lattice which permits atoms to move around their ideal sites within a volume (area) $\Omega$ centered around the lattice sites (bold dots), as shown by the area sprinkled with thin discrete points.
Each atom can move within this area without constraints derived from symmetry, as exemplified by the arrow.}
\label{fig:comm:L}
\end{figure}

Assuming that the magnitude of the distortions is moderate, we may Taylor expand the energy as a function of the atomic positions in the vicinity of the relaxed atomic coordinates up to second order.
This means that we assume that the relaxation energy is mainly due to the pair contributions.\cite{luzw91,sluiter89}
This is visualized in Figure~\ref{fig:comm:L} where each atom is allowed to move (continuously or discretely) within a volume $\Omega$ around its ideal position.\cite{kiku99-02}
Thus, in a lattice gas treatment each site is characterized not only by its occupation variable but also by its displacement vector.
Then the energy of the system per atom can be written as
\begin{equation}
E=\sum_{\alpha \ne ij} v_{\alpha}\xi_{\alpha}+\frac{1}{2 N}\sum_{mn,ij}\iint_{\Omega} V_{mn}(\mathbf{R}_{ij}+\mathbf{u}_i-\mathbf{u}_j)\rho_{mn}(\mathbf{u}_i,\mathbf{u}_j) d \mathbf{u}_i d \mathbf{u}_j ,
\label{equ:chap5:LD}
\end{equation}
where $\alpha$ represents a non-pair cluster, $\mathbf{u}_i$ is the displacement vector of atom $i$ from its ideal site, $\mathbf{R}_{ij}$ is the vector between unrelaxed sites $i$ and $j$, and $N$ is the number of atoms.
The summations of $m,\,n$ in the second term run over atomic species $A$ and $B$, while $ij$ runs over all pairs.
The density matrix $\rho_{mn}(\mathbf{u}_{i},\mathbf{u}_j)$ indicates the probability for an ``$mn$'' type $ij$ atomic pair with site $i$ ($j$) having a displacement $\mathbf{u}_i$($\mathbf{u}_j$).
As for the determination of the pair potentials $V_{mn}$ and the non-pair ECI $v_{\alpha}$, we propose two  practical schemes: the first is based on the lattice inverse
method due to Chen\cite{chen90,chen97,zhang02} and the second employs empirical pair potential models.

\emph{Scheme \uppercase \expandafter{\romannumeral 1}}:
A completely \emph{ab initio} method for finding the pair potential is the lattice inverse method based on the modified M\"{o}bius inverse~\cite{chen90,chen97} from number theory.
It has been shown to give reasonable energy functions for a wide range of materials.\cite{chenmobiuspapers}
First, the cohesive energies $E(a)$ for the pure elements $A$ and $B$ with a simple crystal structure $\beta$, such as fcc or bcc, are computed as a function of the lattice parameter $a$ over a wide range of $a$ values with first-principles methods.
The pair potential between like atoms, say the $AA$ pair, is then extracted from $E(a)$ through an exact transformation\cite{chen90,chen97}
\begin{equation}
V_{AA}^{\beta}(x)=2\sum_{n=1}^{\infty}I^{\beta}(n)E^{\beta}_{A}[b^{\beta}(n)x],
\label{equ:chap5:Vaa}
\end{equation}
where the inverse coefficient $I^{\beta}$ is derived elsewhere~\cite{chen90,chen97,chen98} and its values for fcc and bcc have been conveniently tabulated.\cite{chen97}
The symbol $x$ in Eq.\eqref{equ:chap5:Vaa} is the nearest neighbor distance, while $b^{\beta}(n)$ are coefficients related to the $n^{th}$ coordination shell in the $\beta$ structure, also tabulated elsewhere.\cite{chen90,chen97}
Once the pair potentials $V_{AA}$ and $V_{BB}$ have been obtained, the pair potential related to unlike pairs is determined
from the cohesive energies of some unrelaxed ordered structures with intermediate composition.\cite{chen98}
The cohesive energy due to $AA$ and $BB$ pairs in the ordered structures as computed with the pair potentials derived for the pure elements, is subtracted from the cohesive energy of the ordered structure.
This remainder of the cohesive energy of the ordered structure is due to the unlike bonds and it is inverted just as was done for the pure elements, but of course, now only $AB$ pairs are considered in the structure.\cite{chen98}

In practice, there may be more than one elemental structure $\beta$.
Then, slightly different pair potentials may be generated for the different structures.
Experience shows the potentials derived from different structures to be rather similar,\cite{chenmobiuspapers,GENG1} so that one may opt to average them over the structures $\beta$,
\begin{equation}
V_{AA}(x)=\langle V^{\beta}_{AA}(x)\rangle,\quad \ V_{BB}(x)=\langle V^{\beta}_{BB}(x)\rangle,\quad \ V_{AB}(x)=\langle V^{\beta}_{AB}(x)\rangle .
\label{equ:chap5:Vmm}
\end{equation}

Now that the pair potentials are known, the multi-body, non-pair ECI can be obtained with the conventional CEM procedure performed on the unrelaxed lattice, provided that all the pair potential contributions to the cohesive energy are subtracted out,
\begin{equation}
v_{\alpha}(a)=\sum_{\beta}(\xi^{-1})_{\alpha}^{\beta} \Delta E^{\beta}(a),
\end{equation}
where $(\xi^{-1})_{\alpha}^{\beta}$ is the (pseudo-) inverse of the correlation function matrix involving non-pair clusters $\alpha$ and structure $\beta$.
$\Delta E^{\beta}(a)$ the cohesive energy at a lattice parameter of $a$ of the underlying lattice with the pair potential contributions subtracted out.
$\Delta E^{\beta}(a)$ is derived from the right hand side of Eq.\eqref{equ:chap5:LD},
\begin{align}
\Delta E^{\beta}(a)=E^{\beta}(a) - \frac{1}{2 N} \sum_{ij} \Big[& V_{AA}(\mathbf{R}_{ij})p_{A}^{i}p_{A}^{j}+ V_{BB}(\mathbf{R}_{ij})(1-p_{A}^{i})(1-p_{A}^{j}) \nonumber\\
& + 2 V_{AB}(\mathbf{R}_{ij})p_{A}^{i}(1-p_{A}^{j}) \Big],
\end{align}
where the occupation number $p_{A}^{i}$ takes a value 1 if site $i$ is occupied by the species indicated in the subscript and takes a value 0 otherwise, which relates to the conventional Ising spin-like variable $\sigma$ as $\sigma_{i}=2p^{i}_{A}-1=1-2p^{i}_{B}$.
An interesting aspect of this hybrid cluster expansion is that unrelaxed cohesive energies are needed only - potentially a very significant saving in computational effort.

\emph{Scheme \uppercase \expandafter{\romannumeral 2}}:
Here, the mathematical form of the pair potential is known {\it a priori}, either due to the nature of the bonds or as a matter of expediency.
The $AA$, $BB$, and $AB$ pair potentials are determined through optimization of the adjustable parameters in the pair potential formula.
For simple potentials especially, such as the Lennard-Jones potential,\cite{mohri01} these parameters are found quite simply requiring only a few calculations for unrelaxed structures of the pure elements and unrelaxed ordered structures of intermediate composition, or they could be determined from experimental data.
The multi-body ECI are computedly lastly, in the same manner as under scheme \uppercase \expandafter{\romannumeral 1}.

Initial calculations with the first scheme at zero Kelvin have shown that
NiAl with the unstable $L1_{0}$ structure properly relaxed to the B2 structure with a corresponding energy reduction of about 0.17\ eV per atom and the ratio of the fcc lattice parameters $a/c$ changed from 1 to 1.414 as expected.
In another preliminary test involving CuAu with initially the fcc-based $L1_{0}$ structure, the calculated lattice relaxation resulted in an energy decrease of 0.03\ eV/atom while the $a/c$ ratio became 1.12, to be compared with the experimentally measured value of 1.07.\cite{okamoto87}

In conclusion, a hybrid pair potential - cluster expansion method has been developed which allows an efficient coupling of displacive and substitutional degrees of freedom in alloys.
The main merit of this hybrid cluster expansion is that it allows lattice relaxation to be modeled with relatively minor computational effort.
It provides a more realistic energy functional for use in Monte Carlo, CVM and other lattice gas type simulations for thermodynamic properties of solids at finite temperatures.
Unlike the conventional CEM which employs directly \emph{ab initio} relaxed energies, the hybrid CEM retains
the degrees of freedom associated with relaxation in an explicit form so that relaxations can be extracted from
lattice gas simulations which use the hybrid energy functional.
This means that temperature-dependent relaxations, and processes which are sensitive to local relaxations such as diffusion can be treated more realistically.
Somewhat counter-intuitively, as the hybrid energy functional relies on non-relaxed structural energies it is computationally less demanding to derive than the conventional CEM.

\begin{acknowledgments}
This work was supported by the National Advanced Materials Committee of
China. The authors gratefully acknowledge the financial support from 973
Project in China under Grant No. G2000067101. Part of this work was
performed under the inter-university cooperative research program of the
Laboratory for Advanced Materials, Institute for Materials Research, Tohoku
University.
This work is part of the research program of the Stichting voor Fundamenteel Onderzoek der Materie (Foundation for Fundamental Research of Matter), and was made possible by financial support from the Nederlandse Organisatie voor Wetenschappelijk Onderzoek NWO (Netherlands Organization for Scientific Research).
\end{acknowledgments}


\begin{thebibliography}{99}
\bibitem{GENG1}
For noble and alkali metals it has been confirmed that the pair potentials obtained with the lattice inverse method are rather independent of the particular structure used.
Some minor crystal field effects associated with the directionality of the bonds are present but by averaging over different structures this kind of orientative polarization can be smeared out, restoring the spherical symmetry characteristic of pair potentials.
\end{thebibliography}
\end{document}